\documentclass[iop,12pt]{emulateapj}
\usepackage{natbib} 
\usepackage{amssymb,amsmath}
\usepackage[dvipsnames]{color}
\def\refnew#1{\,(\ref{#1})}
\def\Msun{M_\odot}

\def\ms{{\rm ms}}

\def\Myr{{\rm Myr}}
\def\d {{\rm d}}
\def\yr{{\rm yr}}

\def\Hz{{\rm Hz}}
\def\D{\Delta}

\def\Hc{\mathcal{H}}
\def\J{\mathcal{J}}
\def\I{\mathcal{I}}
\def\B{\mathbf{B}}
\def\yr{{\rm y}}

\def\rin{\mathbf{r}_1}
\def\rout{\mathbf{r}_2}
\def\rj{\mathbf{r}_j}
\def\rinddot{\ddot{\mathbf{r}}_1}

\def\rjdot{\dot{\mathbf{r}}_j}
\def\Jvec{\mathbf{J}}
\def\zhat{\hat{\mathbf{z}}}
\def\pers{\mathrm{s}^{-1}}

\begin{document}

\title{Secular Evolution of Pulsar Triple System J0337+1715}
\author{Jing Luan \& Peter Goldreich}
\email{jingluan@caltech.edu}
\affiliation{California Institute of Technology, Pasadena, CA 91125, US}

{\large \begin{abstract}
The pulsar triple system, J0337+1715, is remarkably regular and highly hierarchical. Secular interactions transfer angular momentum between inner and outer orbits unless their apsidal lines are parallel or anti-parallel. These choices correspond to orthogonal eigenmodes p and a characterized by $e_{p, 1}/e_{p, 2}\sim a_1/a_2$ and $e_{a, 1}/e_{a, 2}\sim (a_1/a_2)^{-3/2}(m_2/m_1)$. 
Mode p dominates the current state so $e_1/e_2$ remains close to $e_{p, 1}/e_{p, 2}$. A small contribution by Mode a causes $e_1$ and $e_2$ to oscillate with period $\sim 10^3\,\yr$ which should be apparent in a few years.  These will reveal the effects of general relativity (GR), and possibly the distortion of the inner white dwarf (WD).
\cite{1992RSPTA.341...39P} proposes that the epicyclic energy of a WD-pulsar binary reaches equipartition with the kinetic energy of a single convective eddy when the WD's progenitor fills its Roche lobe. We extend Phinney's theory to apply to modes rather than individual orbits.  Thus we predict that Mode p and Mode a achieved equipartition with eddies in the giant envelopes of the progenitors of the outer and inner WD, respectively.  The most effective eddies are those with lifetimes closest to the orbit period.  These were far more energetic in the progenitor of the outer WD. This explains why Mode p overwhelms Mode a, and also why the inner binary's orbit is far more eccentric than orbits of other WD-pulsar binaries with similar orbit periods. Mode a's small but finite amplitude places a lower bound of $Q\sim 10^6$ on the tidal quality parameter of the inner WD.
\end{abstract}

\maketitle

\section{Introduction}\label{sec:intro}

PSR J0337+1715 is a $M=1.4\Msun\,\ms$ pulsar with two white dwarf (WD) companions. The inner and outer have masses  
$m_1\simeq 0.2\Msun$ and $m_2\simeq 0.4\Msun$, and move on nearly circular and coplanar orbits with periods of $P_1\simeq 1.6\d$ and $P_2\simeq 327\d$, and eccentricities $e_1\simeq 6.9\times 10^{-4}$ and $e_2\simeq 3.5\times 10^{-2}$. Exquisite timing of the pulsar's pulses enabled \cite{2014Natur.505..520R} to fit the system's parameters to striking accuracy.   Because of the large disparity in orbital periods, mean motion resonances may be neglected.  Thus secular interactions govern the system's long-term evolution. 

Secular dynamics is an approximation in which each body is replaced by an elliptical wire spread along its orbit. Unless apsidal lines align, mutual torques give rise to angular momentum exchanges. Fractional energy exchanges are much slower than those of angular momentum, $|\dot E/E|\ll |\dot J/J|$, where $E$ and $J$ denote orbital energy and angular momentum.  We ignore energy exchanges and take semi-major axes to be constants.  

Conventional celestial mechanics was developed to study the solar system. It is optimized for systems with a massive central body and multiple much smaller ones.  Two-body Keplerian orbits, each consisting of the central mass and one of its companions, comprise the zeroth order state.  Perturbations due to gravitational interactions among the small bodies modify the Keplerian orbital elements. Because the WD masses are within an order of magnitude of the mass of the neutron star (NS), the standard procedure fails for the pulsar triple system under consideration here .  A more appropriate procedure is to take the unperturbed system to consist of an inner and an outer binary with the former made up of the NS and the $0.2\Msun$ WD and the latter by the $0.4\Msun$ WD and the inner binary.  Perturbations due to interactions between the binaries are small because their semi-major axes are so different \citep{2000ApJ...535..385F, 2014arXiv1403.5292R}.  

Our paper is organized as follows.  We develop a compact formalism for secular interactions in a hierarchical coplanar triple system in Section \ref{sec:secular-formalism}. Included are precession terms due to general relativity (GR) along with tidal and rotational distortions of the inner WD. Interactions of modes with convective eddies are considered in Section \ref{sec:mode-damp}.  Section \ref{sec:secular-oscillation} applies our model to J0337+1715. We predict that secular evolution will be detected in the near future and that the effects of general relativity (GR) will become apparent. In Section \ref{sec:compare}, we compare our secular formalism with results obtained by direct numerical integrations obtained using the program Mercury \citep{1999MNRAS.304..793C}.  The agreement is good except that the numerical results yield short-term variations of the osculating eccentricities.  We argue that these are deceptive. Section \ref{sec:conclusion} summarizes our conclusions. 

Although couched in different language, some material in our paper is closely related to that in \cite{2000ApJ...535..385F} and \cite{2014arXiv1403.5292R}. The former provides a more general treatment of hierarchical triple systems than we do whereas the latter contains an analytic analysis of the same system that we are investigating. In areas of overlap, our results agree with those obtained in \cite{2014arXiv1403.5292R}.

\section{Secular Interaction in Co-planar Triple System}\label{sec:secular-formalism}

We use 1 and 2 to label the inner and outer binary orbits and/or WDs. 
We denote the vector from the NS to the inner WD by $\rin$ and the one from the center of mass of the inner binary to the outer WD by $\rout$.  Manipulation of the inertial-frame equations of motion yields
\begin{eqnarray}
\mu_1\rinddot &=&-\nabla_{1} U \, ,\\
\mu_2\ddot{\rout} &=& -\nabla_{2} U \, ,
\end{eqnarray}
with potential
\begin{equation}\label{eq:U}
U=- {G M m_1\over r_1}-{G M m_2\over \left| {m_1\over M+m_1}\rin +\rout \right|}-{G m_1 m_2 \over \left| {M\over M+m_1} \rin-\rout \right|}\, .
\end{equation}
Here $\nabla_j$ indicates gradient with respect to $\rj \, (j=1,2)$ and reduced masses are $\mu_1\equiv M m_1/(M+m_1)$ and $\mu_2\equiv (M+m_1)m_2/(M+m_1+m_2)$. We average $U$ over the zeroth-order orbits described by $\rin=r_1(\cos\theta_1,\sin\theta_1,0)$, $\rout=r_2(\cos\theta_2,\sin\theta_2,0)$ and 
\begin{eqnarray}
r_1&=& {a_1(1-e_1^2)\over 1+e_1\cos(\theta_1-\varpi_1)} \, ,\\
r_1^2 \dot{\theta}_1 &=& \left(G(M+m_1)a_1(1-e_1^2)\right)^{1/2} \, ,\\
r_2 &=& {a_2(1-e_2^2)\over 1+e_2\cos(\theta_2-\varpi_2)} \, ,\\
r_2^2\dot{\theta}_2 &=& \left(G(M+m_1+m_2)a_2(1-e_2^2)\right)^{1/2} \, ,
\end{eqnarray}
where $a$ is semi-major axis, $e$ is eccentricity, and $\varpi$ is the longitude of pericenter. 
The mean motions are $n_1\equiv (G(M+m_1)/a_1^3)^{1/2}$ and $n_2\equiv (G(M+m_1+m_2)/a_2^3)^{1/2}$.
Retaining terms up to second order in eccentricity,\footnote{The zeroth order term is constant, and the one of first order vanishes.} yields the secular potential 
\begin{eqnarray}
U_{\sec}&=& -{3\over 8}\alpha^2 (e_1^2+e_2^2) {G M m_1 m_2\over a_2 (M+m_1)}+{15\over 16}\alpha^3 (e_1 e_2)\nonumber \\
&& \times \cos(\varpi_1-\varpi_2){G M m_1 m_2(M-m_1)\over a_2 (M+m_1)^2} \, ,
\label{eq:Usec}
\end{eqnarray}
where $\alpha\equiv a_1/a_2$.

We verify that the inertial-frame total angular momentum is given by
\begin{eqnarray}
\Jvec&=& \sum_{j=1}^2 \Jvec_j\equiv \sum_{j=1}^2 \mu_j \rj \times \rjdot \nonumber\\
&=&\zhat \left(\sum_{j=1}^{2} J_{j,c}-\delta J_j\right) \, , \label{Jtot}
\end{eqnarray}
where $\zhat =(0,0,1)$. The circular part of angular momentum for each orbit, $J_{j,c}\equiv \mu_j a_j^2 n_j$, is conserved because $a_j$ is constant. It is the angular momentum deficit, $\delta J_j\equiv J_{j,c} e_j^2/2$, that is exchanged under interactions. The total angular momentum deficit reads
\begin{equation}
\J = \sum_{j=1}^{2} \delta J_j = \I^{\dagger}\I \, ,
\end{equation}
where $\I$ is a two dimensional column vector with components $I_j\equiv (\mu_j n_j/2)^{1/2} a_j e_j\exp(i\varpi_j)$ and $\dagger$ denotes Hermitian conjugate.  

Secular evolution is governed by the equation of motion
\begin{equation}\label{eom}
\dot \I=i \B \I \, .
\end{equation}
The Hamiltonian $\Hc\equiv U_{\rm sec}$ reads
\begin{eqnarray}\label{eq:Ham}
\Hc&=&  - \I^{\dagger}\B\I \, .
\end{eqnarray}
Conservation of both $\Hc$ and $\J$ follows immediately from equations (\ref{eom}) and (\ref{eq:Ham}).  Elements of the symmetric $2\times 2$ matrix
\begin{equation}
\B\equiv
\left(\begin{array}{cc}
B_{11} & B_{12}\\
B_{12} & B_{22}
\end{array}
\right)\label{eq:matrixB}
\end{equation}
read
\begin{eqnarray}
B_{11}&=& {3\over 4}n_1\alpha^3{m_2\over M+m_1}\, ,\label{B11}\\
B_{22}&=& {3\over 4} n_1 \alpha^{7/2}{M m_1(M+m_1+m_2)^{1/2}\over (M+m_1)^{5/2}}\, ,\label{B22} \\
B_{12}&=&-{15\over 16}n_1 \alpha^{17/4}{(M-m_1)(Mm_1m_2)^{1/2}\over (M+m_1)^{11/4}}\nonumber \\
&& \times (M+m_1+m_2)^{1/4}\, .
\end{eqnarray} 
{\cite{2000ApJ...535..385F} previously studied secular theory for highly hierarchical triple systems. 
Our $\dot e$ and $\dot \varpi$ are derivable from their equations (46)-(52). 

Substituting the trial solution, $\I\propto \exp(i gt)$, into Eq.\refnew{eom}, we obtain two modes represented by the eigenvalues and normalized eigenvectors of $\B$. Mode p (parallel) has $\varpi_1=\varpi_2$, and Mode a (anti-parallel) has $\varpi_1=\varpi_2+\pi$.  In each mode, both pericenters precess at the same rate which is the corresponding eigenvalue,
\begin{eqnarray}
g_p&=& {1\over2}\left((B_{11}+B_{22})-\D g \right)\, ,\label{gp}\\
g_a&=& {1\over 2}\left((B_{11}+B_{22})+\D g \right)\, .\label{ga} 
\end{eqnarray}
The relative precession rate
\begin{equation}
\D g \equiv g_a-g_p=\sqrt{(B_{11}-B_{22})^2+4B_{12}^2}\, .\label{dg}
\end{equation}
Components of the eigenvectors satisfy
\begin{eqnarray}
\frac{I_{p, 1}}{I_{p, 2}}= -\frac{I_{a, 2}}{I_{a, 1}}=\frac{\left(-B_{11}+B_{22}+\D g \right)}{2|B_{12}|}\, .
\label{I1I2ratio} 
\end{eqnarray}
The first minus sign in Eq.\refnew{I1I2ratio} appears because $\exp i(\varpi_1-\varpi_2)=-1$ for Mode a. The general solution is a linear combination of the two modes,
\begin{equation}\label{I-t}
\I(t)=c_p\hat\I_p\exp(i g_p t)+c_a\hat\I_a\exp(i g_a t)\, ,
\end{equation}
where ${\hat {}}$  signifies a normalized vector. Unless either $c_p$ or $c_a$ vanishes, $e_1/e_2$ and $\varpi_1-\varpi_2$ oscillate at frequency $\D g$. 

Since $\B$ is symmetric, $\hat\I_a^{\dagger}\hat\I_p=0$. Contributions from $\hat\I_a$ and $\hat\I_p$ to $\Hc$ and $\J$ are thus separable,
\begin{eqnarray}
\Hc &=&{1\over 2} g_p |c_p|^2+{1\over 2} g_a |c_a|^2\equiv \Hc_p+\Hc_a \, ,\nonumber\\
\J&=& {1\over 2} |c_p|^2+{1\over 2} |c_a|^2\equiv \J_p+\J_a\, .\label{H-J-separate}
\label{eq:actions}
\end{eqnarray}
The ratios $\Hc_a/\Hc_p$ and $\J_a/\J_p$ are measures of the relative strengths of Modes a and p.
We note that $|c_p|^2$ and $|c_a|^2$ are actions and thus might behave as adiabatic
invariants under slow variations of the masses and orbits.  

Given that $\alpha\ll 1$ and $B_{11}\gg B_{22}\gg B_{12}$, the following approximations apply:
\begin{eqnarray}
g_p &\sim& B_{22} \ll g_a \sim B_{11}\, ,\\
\D g&\sim & B_{11}-B_{22}\, .\label{dg-approx}
\end{eqnarray}
Thus Mode p precesses a factor $g_p/g_a\sim \alpha^{1/2}(m_1/m_2)$ more slowly than Mode a.  For the eigenvectors,
\begin{equation}
\frac{I_{p, 1}}{I_{p, 2}}= -\frac{I_{a, 2}}{I_{a, 1}}\sim {|B_{12}|\over B_{11}} \sim \alpha^{5/4}\left(m_1\over m_2\right)^{1/2}\, ,
\end{equation}
which yields
\begin{eqnarray}
\frac{e_{p, 1}}{e_{p, 2}}&\sim& \alpha\, ,\\
\frac{e_{a, 1}}{e_{a, 2}}&\simeq& \alpha^{-3/2} \frac{m_2}{m_1}\, .
\end{eqnarray}

\subsection{Additional Precession Rates}\label{subsec:additional-precession}
General relativity (GR) and tidal and rotational distortions of the inner WD cause its pericenter to precess forward at rates \citep{2002ApJ...564.1024W},
\begin{eqnarray}
\left.\frac{d\varpi_1}{dt}\right|_{GR}\ &=& 3 n_1 \frac{G(M+m_1)}{c^2 a_1}\, ,\label{w1dotgr} \\
\left.\frac{d\varpi_1}{dt}\right|_{\mathrm{tide}}&=& {15\over 2} n_1 k_2 \frac{M}{m_1}\left(\frac{R_1}{a_1}\right)^5\, ,\label{w1dottide} \\
\left.\frac{d\varpi_1}{dt}\right|_{J_2}\ \ &=& {1\over 2} n_1 k_2 \left(\frac{\Omega_1}{n_1}\right)^2 \frac{M}{m_1}\left(\frac{R_1}{a_1}\right)^5\, .\label{w1dotJ2}
\end{eqnarray}
The Love number $k_2\simeq 0.29$ for an $n=1.5$ polytrope (\cite{1933MNRAS..93..449C}), which is a reasonable proxy for a WD.  The WD's radius is denoted by $R$. We neglect these effects on the outer WD because $a_2\gg a_1$ and all the above precession rates decline with distance from the NS. The pulsar is so dense that its tidal and rotational deformations are negligible. 
We denote the total additional precession rate of $\dot\varpi_1$ by $\Delta$ and add it to $B_{11}$. In so doing, we make the plausible assumption that inner WD's spin speed, $\Omega_1$, is synchronized with its mean motion, $n_1$.

\section{Mode Damping and Excitation}\label{sec:mode-damp}

\subsection{Damping}\label{subsec:damp}
Dissipation associated with the tides raised in the WDs by the pulsar act to damp their orbital eccentricities.  We define $\tau_j\equiv e_j/\dot e_j\vert_{\mathrm{damp}}$, which implies a complementary change rate for $I_j$, namely $\dot I_j\vert_{\mathrm{damp}}=-I_j/\tau_j$ which we account for by replacing $B_{jj}$ by $B_{jj}+i/\tau_j$ in Eq.\refnew{eq:matrixB}. Because $\tau$'s are much longer than the precession period, they introduce small corrections to $\B$. Thus we retain the old eigenvectors while expanding the eigenvalues to first order in the $1/\tau_j$.  This procedure adds damping terms to $g_p$ and $g_a$ which read:
\begin{eqnarray}
\gamma_p&=& {1\over 2\tau_2}\left(1+{B_{11}-B_{22}\over \D g}\right)+{1\over 2\tau_1}\left(1-{B_{11}-B_{22}\over \D g}\right)\nonumber\\
&\simeq & {1\over \tau_2}+\frac{m_1}{m_2} \frac{\alpha^{5/2}}{\tau_1}\, \label{eq:gammap}
\end{eqnarray}
and
\begin{eqnarray}
\gamma_a&=& {1\over 2\tau_1}\left(1+{B_{11}-B_{22}\over \D g}\right)+ {1\over 2\tau_2}\left(1-{B_{11}-B_{22}\over \D g}\right)\nonumber\\
&\simeq & {1\over \tau_1}+\frac{m_1}{m_2} \frac{\alpha^{5/2}}{\tau_2
}\, \label{eq:gammaa}
\end{eqnarray}
As a consequence of the orbital eccentricity ratios in modes p and a, dissipation in the inner WD selectively damps Mode a and that in the outer WD selectively damps Mode p.  

\subsection{Excitation}\label{subsec:excite}
\cite{1992RSPTA.341...39P} argues that the orbital eccentricity of a binary composed of a pulsar and a low-mass WD is set during the final stages of Roche lobe overflow (RLO) by the WD's progenitor.\footnote{During RLO, these systems are observed as low mass x-ray binaries (LMXRB)}  Convection in the progenitor's extended envelope creates a fluctuating quadrupole that stochastically excites orbital eccentricity while turbulent viscosity simultaneously damps it. These competing processes drive the epicyclic energy, $E_{\mathrm ec}\equiv n\delta J$, toward equipartition with the kinetic energy of eddies whose lifetimes are closest to the orbit period. Equipartition is approached on the eccentricity damping timescale, $\tau_e$, which is much shorter than the duration of RLO. Eccentricities established in this manner increase with orbit period as a consequence of the increase in eddy kinetic energy with eddy lifetime.  Observational data offers support for Phinney's proposal \citep{2012MNRAS.425.1601T}.  In what follows, we apply the equipartition concept to the pulsar triple system. However, our focus is on the epicyclic energies of modes p and a rather than those of the binary orbits.  

Modal epicyclic energies are defined by
\begin{eqnarray}
E_{p, \mathrm ec}&\equiv & n_1|I_{p, 1}|^2+n_2|I_{p, 2}|^2\,\cr 
&\equiv& \frac{1}{2}\mu_1(n_1a_1e_{p, 1})^2+\frac{1}{2}\mu_2(n_2a_2e_{p, 2})^2\,  
\label{eq:Epec}
\end{eqnarray}
and 
\begin{eqnarray}
E_{a, \mathrm ec}&\equiv& n_1|I_{a, 1}|^2+n_2|I_{a, 2}|^2\,\cr 
&\equiv& \frac{1}{2}\mu_1(n_1a_1e_{a, 1})^2+\frac{1}{2}\mu_2(n_2a_2e_{a, 2})^2\,  .
\label{eq:Eaec}
\end{eqnarray}
These epicyclic energies are constants of motion under evolution governed by the Hamiltonian in equation \refnew{eq:Ham}.   Epicyclic energies may also be defined for the inner and outer binary orbits.  They vary on the secular timescale.  Moreover, their sum also varies and only equals the sum of the modal epicyclic energies when the inner and outer binary apses are either parallel or anti-parallel, $\sin(\varpi_1-\varpi_2)=0$. 

Tidal interactions during the LMXB stage that gave rise to the outer WD drove the epicyclic energy of Mode p toward equipartition with a single eddy. Based on estimates for the duration of Roche lobe overflow and the rate of eccentricity damping by tides, energy equipartition should have persisted until termination of the LMXB phase.  During the outer LMXB phase, $\tau_1$ is essentially infinite. Thus according to Eqs.\refnew{eq:gammap} and \refnew{eq:gammaa}, $\gamma_a\sim  10^{-4}\gamma_p$. It follows that the epicyclic energy of Mode a probably experienced little progress toward equipartition.  The same story applied during the inner LMXB stage, except the roles of Mode p and Mode a were reversed.  
In this scenario, the order in which the inner and outer WDs formed is not crucial.  In either case, the epicyclic energy in Mode p ends up much larger than that in Mode a because the energies of convective eddies whose lifetimes are closest to the orbital period are greater for larger orbital periods. 

As mentioned previously, the modal angular momentum deficits, $\J_p$ and $\J_a$ are actions of the Hamiltonian given by equation \refnew{eq:Ham}.  Thus they are invariants under slow changes of masses and semi-major axis provided these occur independently of the secular oscillation.  For example, orbital evolution during the formation of the second WD would change the epicyclic energy of the mode which achieved equipartition during the birth of the first WD.  If the angular momentum deficit of the mode remained invariant, its epicyclic energy would change in proportion to the change of its precession rate.


\section{Secular Oscillation of J0337+1715}\label{sec:secular-oscillation}
We adopt stellar masses, orbital periods, eccentricities and longitudes of pericenter from Table 1 in \cite{2014Natur.505..520R}. The semi-major axes are obtained from $a_1= (G(M+m_1)/n_1^2)^{1/3}$ and $a_2=(G(M+m_1+m_2)/n_2^2)^{1/3}$. 
Inserting these parameters into the elements of matrix $\B$, we obtain 
\begin{equation}
\B\simeq \left(\begin{array}{cc}
2.34405 & -0.0119143 \\
-0.0119143 & 0.129615
\end{array}\right)\times 10^{-10}\,\pers \, .
\end{equation}
Here $B_{11}$ includes $\D$, 
\begin{eqnarray}
\Delta&\equiv& \left.\frac{d\varpi_1}{dt}\right|_{\mathrm{GR}}+\left.\frac{d\varpi_1}{dt}\right|_{\mathrm{tide}}+\left.\frac{d\varpi_1}{dt}\right|_{\mathrm{J_2}}\nonumber\\
&& \nonumber\\
&\simeq& (6.77+0.0315+0.00210)\times 10^{-11}\, \pers\nonumber\\
&\simeq& 6.80\times 10^{-11}\,\pers\, ,\label{D-numeric}
\end{eqnarray}
which is dominated by GR and contributes $\simeq 30\%$ to $B_{11}$.
Table \refnew{Compare-D} and Fig.\refnew{e-w-t} compare results obtained by either excluding or including $\D$ from which we make several observations:
\begin{deluxetable}{ccccc}
\tabletypesize{\scriptsize}
\tablecaption{Secular Oscillation\label{Compare-D}}
\tablewidth{0pt}
\tablehead{
\colhead{} & \colhead{Parameter} & \colhead{Without $\D$} & \colhead{With $\D$} & \colhead{Difference} 
}
\startdata
1& $g_p\ (10^{-11}\Hz)$ & $1.2952$ & $1.2955$ & $-2.8$E-4 \\   
2 & $g_a\ (10^{-11}\Hz)$ & 16.6 & 23.4 & -6.8 \\ 
3 & $\D g\ (10^{-11}\Hz)$ & 15.3 & 22.1 & -6.8\\ 
4 & $2\pi /\D g\ (\mathrm{kyr})$ & 1.298 & 0.899 & 0.40\\ 
5 & $(e_1/e_2)_p$ & 0.0278 & 0.0193 &  $8.5$E-3\\
6 & $(e_1/e_2)_a$ & 461.4 & 665.9 & -204.6 \\
7 & $\J_a/\J_p/10^{-8}$ & $536.6$ & $4.5$ & 532.1  \\
8 & $E_{a,\mathrm ec}/E_{p, \mathrm ec}/10^{-6}$ & $64.9$ & $1.01$ & 63.9 \\ 
9 &$\delta(e_1\cos\varpi_1)/10^{-8}$ & $114.0$ & $-33.06$  & $147.1$  \\
10 & $\delta(e_1\sin\varpi_1)/10^{-8}$ & $-1.620$ & $ -21.94$ & $ 20.32$  \\
11 & $\delta(e_2\cos\varpi_2)/10^{-9}$ & $-1.44$E4 & $-1.44$E4 & $1.1$E-3  \\
12 & $\delta(e_2\sin\varpi_2)/10^{-9}$ & $-1.42$E3 & $-1.42$E3 & $-7.7$E-3 \\
\enddata
\tablecomments{Columns 3 and 4 show results without and with $\D$ included in $B_{11}$. Column 5 is calculated by subtracting Column 4 from Column 3. The expression Ex means $10^x$ in some of the numerical values. According to \cite{2014Natur.505..520R}, the accuracies of measurements for $e_1\cos\varpi_1$ and $e_1\sin\varpi_1$ are $10^{-8}$, and those for $e_2\cos\varpi_2$ and $e_2\sin\varpi_2$ are $10^{-9}$. Rows 9 to 12 show their changes after $1\yr$, in multiples of their corresponding accuracies. Although the secular oscillation occurs on a thousand-year timescale, 
changes over $1\yr$ already exceed measurement accuracies. Differences between predictions without and with $\D$ are also detectable in $\delta(e_1\cos\varpi_1)$ and $\delta(e_1\sin\varpi_1)$.  These could help test GR and constrain the tidal Love number, $k_2$, of the inner WD.}
\end{deluxetable}

\bigskip


\begin{enumerate}

\item According to rows 7 and 8, the system is currently dominated by Mode p, which is consistent with the equipartition scenario assumed in Section \ref{sec:mode-damp}. 

\item Row 4 shows that $\D$ induces a shorter secular oscillation period, as predicted by Eq.\refnew{dg-approx} since $\D$ enhances $B_{11}$. It also shrinks the oscillation amplitudes by an order of magnitude as exhibited by comparing
Fig.\refnew{compare-secular-Mercury} with the solid line in the upper panel of Fig.\refnew{e-w-t}. Secular oscillations appear because the system's state is not pure Mode p, but also contains a small contribution from Mode a. 

\item We see, from rows 9-12, that changes over a year exceed the measurement accuracies cited by \cite{2014Natur.505..520R}.  Especially, measurements of $e_1\cos\varpi_1$ and $e_1\sin\varpi_1$ can separate the cases with and without $\D$. If an accurate determination of $\D$ were made, it would constrain $k_2$ of the inner WD (cf. Eqs. \ref{w1dotgr}-\ref{w1dottide}).
 
\item From private communications with Scott M. Ransom and Anne M. Archibald, we know that in fitting their timing data, orbit elements were assumed constant and GR was not taken into account.  With additional data,  both restrictions may be lifted.  As described in the previous paragraph, the effects of GR should be apparent within a short time.  

\end{enumerate}

\section{Comparison with Numerical Integration}\label{sec:compare}
We ran Mercury \citep{1999MNRAS.304..793C}, a symplectic integrator for Newtonian orbital dynamics, to evolve the pulsar triple system over a secular period $\sim 10^3\yr$. Secular changes of $e_1$ were extracted from a 10 year average of the osculating orbital elements.  As displayed in Fig.\refnew{compare-secular-Mercury}, the numerical and analytical results compare well; oscillation periods differ by $\simeq 0.6\%$ and oscillation amplitudes differ by $\simeq 2\%$.  These differences are reduced by roughly half if the secular Hamiltonian is expanded to fourth order in both $\alpha$ and eccentricities. Inclusion of even higher order terms yields negligible improvement.  

\begin{figure}
\includegraphics[width=0.9\linewidth]{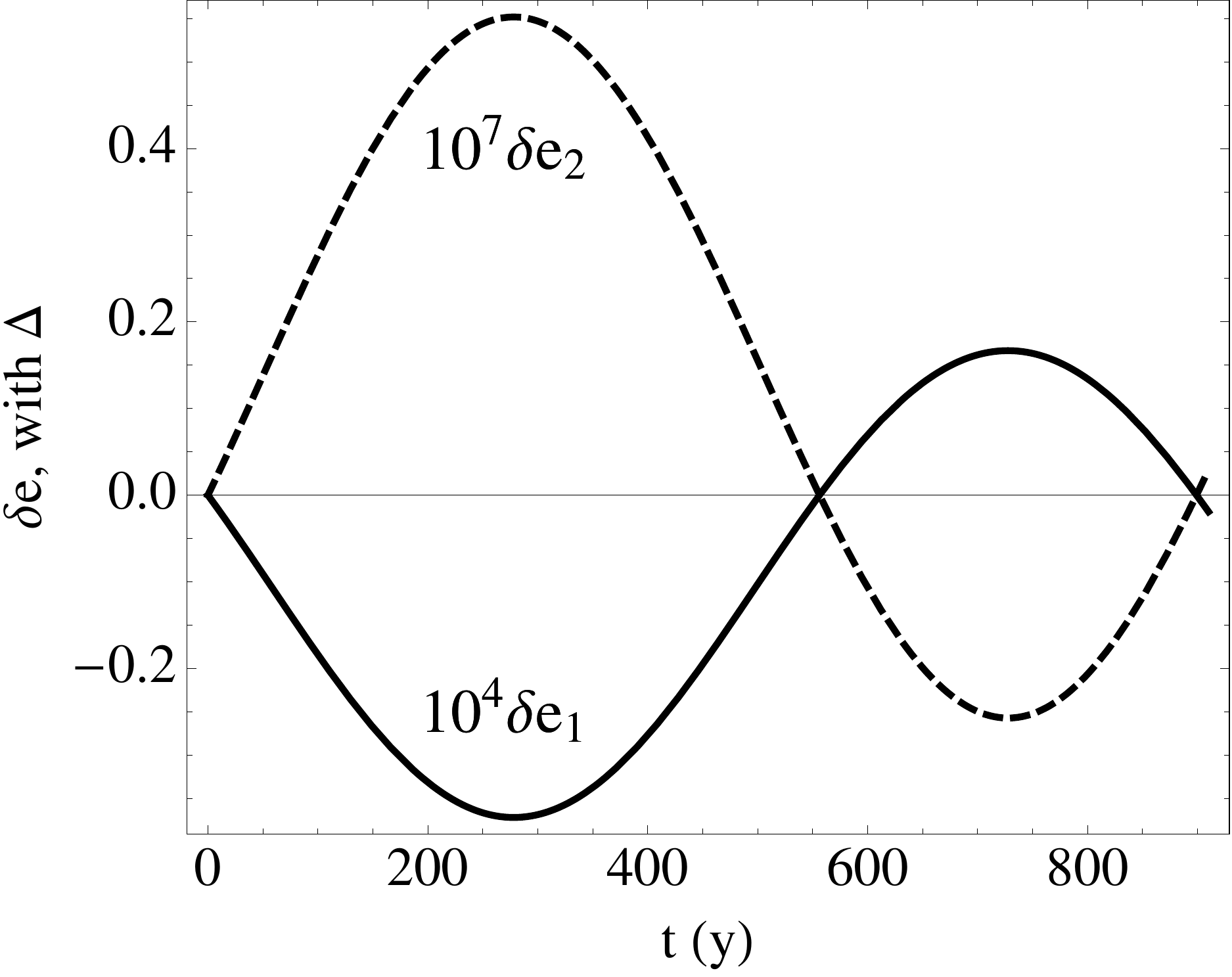}
\includegraphics[width=0.9\linewidth]{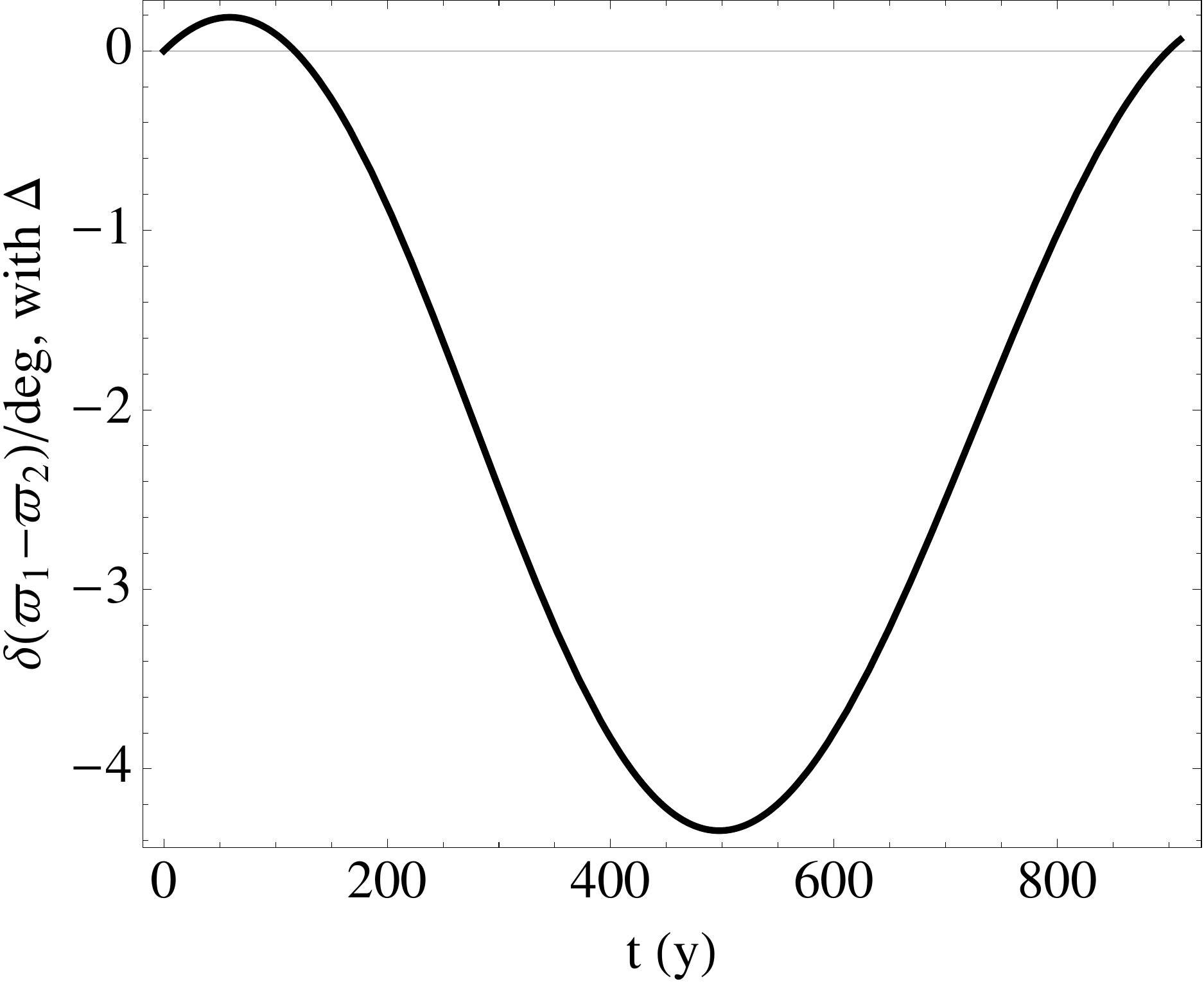}
\caption{\label{e-w-t} Upper panel: solid line: $10^4 \delta e_1$; dashed line: $10^7\delta e_2$.  Lower panel: solid line: $\delta (\varpi_1-\varpi_2)$ in degrees. }
\end{figure}

\begin{figure}
\includegraphics[width=0.9\linewidth]{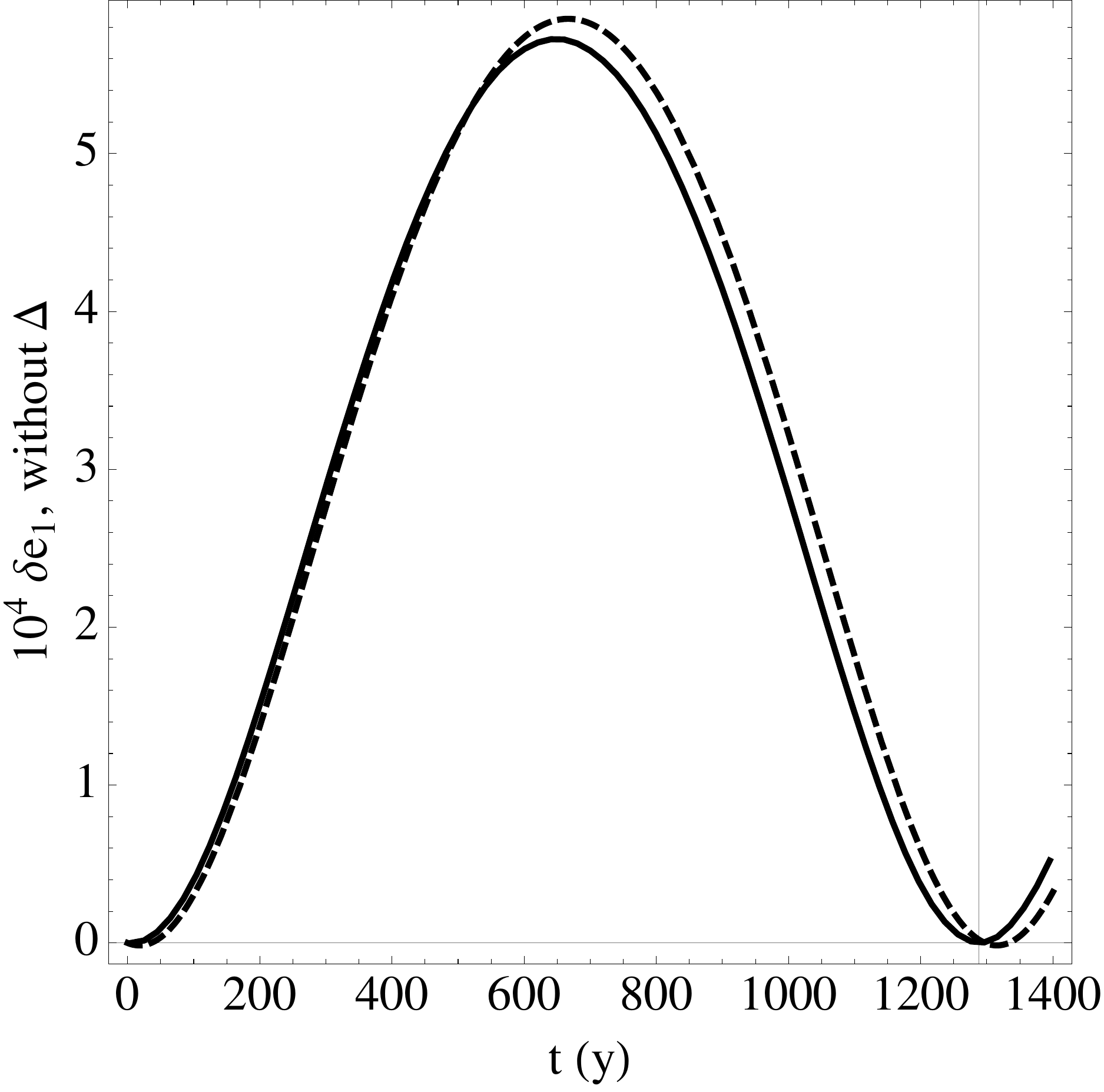}
\caption{\label{compare-secular-Mercury}Comparison of the secular variations of $e_1$ over a precession cycle obtained from a numerical integration with Mercury (solid line) and the analytical model (dashed line) described in Section \ref{sec:secular-formalism}  with $\D=0$.  Oscillation periods are $1289\yr$ and $1297\yr$ respectively.} 
\end{figure}

\subsection{Osculating Elements are Deceptive}\label{subsec:deception}

\begin{figure}
\includegraphics[width=0.9\linewidth]{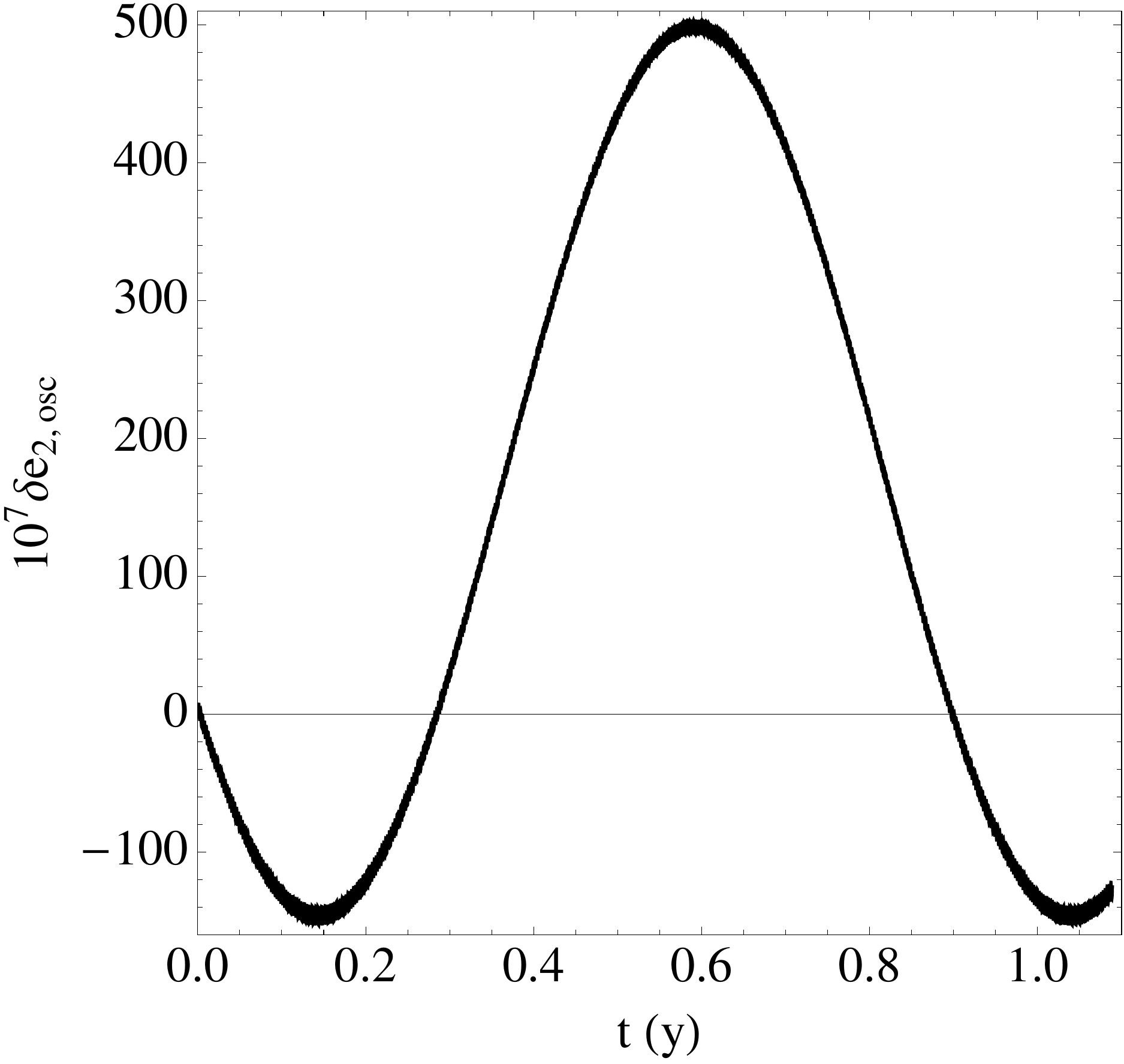}
\caption{\label{fig:de2-P2} The variation of $e_2$ on the outer orbital period.}
\end{figure}

\begin{figure}
\includegraphics[width=0.9\linewidth]{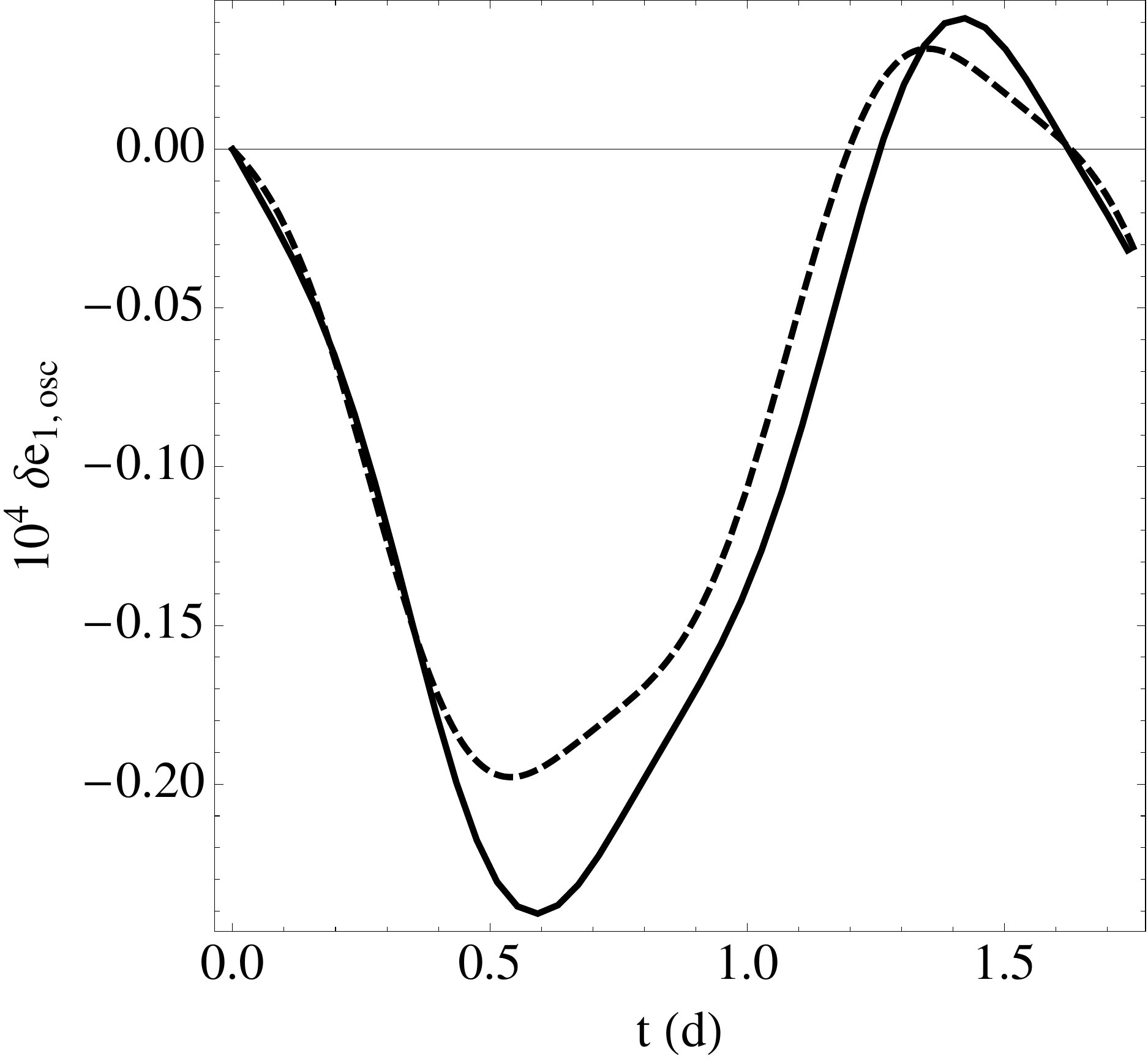}
\caption{\label{fig:de1-P1} The variation of $e_1$ on the inner orbital period. Solid: $\delta e_1$ from the output from Mercury's simulation. Dashed: $\delta e_1$ from analytic solution for the inner orbit with the outer WD's quadruple potential.  }
\end{figure}

Classical perturbation theory is formulated in terms of osculating orbit elements.  These are obtained by fitting a Keplerian ellipse to the instantaneous position and velocity of a perturbed orbit. Temporal variations of the elements describe how the perturbations evolve.  Results of numerical integrations of N-body systems such as Mercury are often expressed in terms of oscillating elements.  Osculating eccentricities obtained from short intervals of output from Mercury are displayed in figures \ref{fig:de2-P2} and \ref{fig:de1-P1}.\footnote{This is a purely Newtonian simulation. GR is not included.}  Each varies over the corresponding orbit period.\footnote{In this subsection, we neglect the small difference between orbit and epicyclic frequencies.} These oscillations are deceptive. In the following paragraphs we explain how they arise. Our focus is on the inner binary orbit because the variation of its osculating eccentricity is more complex.

As a first step, consider a circular orbit in a potential consisting of a dominant monopole and an external axisymmetric quadrupole.  We define $\beta\ll 1$ to be the ratio of the quadrupole to monopole potential evaluated at the orbit's radius, $r_0$.  A simple exercise shows that the corresponding osculating elements $a_{\mathrm osc}\approx (1+\beta)r_0$ and $e_{\mathrm osc}\approx \beta$. Paradoxically, the osculating eccentricity does not vanish.  Moreover, the true anomaly stays fixed at $f_{\mathrm osc}=0$ while the apse rotates with angular velocity $\dot{\varpi}_{\mathrm osc}\approx n$. 

Next we consider a less trivial example in which the orbit possesses a real eccentricity $e$.  However,  we maintain the assumption of an axisymmetric quadrupole. Here we obtain $e_{\mathrm osc}^2\approx e^2+2\beta e\cos(nt+\phi)+\beta^2$ along  with the same $a_{\mathrm osc}\approx (1+\beta)r_0$ as before.  The value of $\phi$ is set as an initial condition.  In the limit $\beta\ll e$, $e_{\mathrm osc}$ oscillates harmonically about $e$ with amplitude $\beta$. Figure \ref{fig:de2-P2} displays an example of this behavior. Because $P_{\mathrm{orb,1}}\ll P_{\mathrm{orb,2}}$, the effects of the non-axisymmetric part of the quadruple potential almost average to zero. 

Lastly, we keep all the quadrupole terms depending on $r_1$ in equation \refnew{eq:U}, which reduces to
\begin{equation}
U=-{G M m_1\over r_1}-{G m_2 \mu_1\over r_2}\left(r_1\over r_2\right)^2 P_2(\cos\D\theta)\, ,
\end{equation}
$\D\theta=\cos^{-1}({\hat {\mathbf r}}_1.{\hat {\mathbf r}}_2)$ circulates at frequency $\sim n_1$. Thus $P_2(\cos\D\theta)\sim(1+3\cos(2nt))/4$.  As a consequence, the epicyclic motion possesses a forced oscillation at frequency $2n_1$ in addition to its free oscillation at frequency $n_1$.  These two terms, together with the angle between the outer body's longitude and the inner body's apse, are responsible for the non-harmonic behavior illustrated by figure \ref{fig:de1-P1} which compares $e_{1,\mathrm osc}$ derived analytically with the result obtained from Mercury.

\section{Discussion and Conclusions}\label{sec:conclusion}
We study secular interactions in the pulsar triple system J0337+1715 utilizing a simplified version of a formalism for highly hierarchical triple systems developed by \cite{2000ApJ...535..385F}.  To second order in the orbital eccentricities, the secular evolution is described in terms of two orthogonal modes.    
In Mode p, the apses of the inner and outer binary orbit align whereas they are anti-aligned in Mode a. 
Mode a precesses more rapidly than Mode p.  Eccentricities of the binary orbits oscillate at the secular frequency, the difference between the precession frequencies of Mode a and Mode p.  The secular frequency corresponds to a long timescale $\sim 10^3\yr$. Nevertheless, secular changes are potentially detectable in the near future, thanks to the exquisite accuracy of measurements by \cite{2014Natur.505..520R}. 
These should easily reveal the effects of GR on both the period and amplitude of the secular oscillation. Although more challenging, it might be possible to detect similar effects from the tidal and rotational deformations of the inner WD and thereby constrain its Love number. 
We generalize the beautiful theory of \cite{1992RSPTA.341...39P}  to apply to the excitation of eigenmodes.  In this form it explains why the current system is dominated by Mode p with Mode a making only a minor contribution even to the eccentricity of the inner binary's orbit.

Tidal dissipation in the inner WD mainly damps Mode a.  The mode's current amplitude, although small with respect to that of Mode p, is at the high end of what might be expected from the eccentricities of pulsar He-core WD binaries with orbital periods of order a few days. Thus it is unlikely that Mode a's tidal damping timescale is much shorter than $\sim 500\,\Myr$, the age we estimate from WD cooling models in \cite{2013A&A...557A..19A}.  This enables us to place a lower limit on the effective tidal $Q$ parameter for this particular WD during its lifetime of
\begin{equation}
Q\simeq ((1+k_2)^2 M^2 n_1 R_1^8 \tau)/(a_1^8 m_1^2) \ga 10^6 \, .
\end{equation}
By comparison, \cite{2011ApJ...740L..53P} assumes that luminosities in the binary WD system J0651 are contributed by asynchronous tidal heating and thereby sets upper limits for $Q$ of $\sim 7\times 10^{10}$ for the He-core WD and $\sim 2\times 10^7$ for the CO-core WD. 

\section*{acknowledgements}

We thank Scott Ransom and Anne Archibald for educating us about how the masses and orbits of the pulsar triple system were derived. We appreciate that Scott Ransom pointed out that we used a wrong number for this system immediately after we posted the paper on arxiv.  
We are grateful to John Chambers both for making his code Mercury publicly available and for responding to our queries on how to use it.  Thanks are also due to Christian Ott and Sterl Phinney for guidance regarding the
possibility that the pulsar in J0337+1715 formed by accretion induced collapse of a WD. We also like to thank Anthony L. Prio for reminding us that the $Q$ values in \cite{2011ApJ...740L..53P} are actually upper limits.


\end{document}